\documentclass[11pt,a4paper]{amsart}
\usepackage{amsfonts,amsmath,amssymb,amsthm}
\usepackage{times}
\usepackage{color}
\usepackage{esint}
\usepackage{pstricks}

\numberwithin{equation}{section}

\renewcommand{\epsilon}{\varepsilon}

\DeclareSymbolFont{SY}{U}{psy}{m}{n}
\DeclareMathSymbol{\emptyset}{\mathord}{SY}{'306}

\DeclareMathOperator*{\slim}{s-lim}
\DeclareMathOperator*{\wlim}{w-lim}

\DeclareMathOperator{\Ran}{Ran} \DeclareMathOperator{\Ker}{Ker}
 \DeclareMathOperator{\Dom}{Dom}
\DeclareMathOperator{\sign}{sign} \DeclareMathOperator{\grad}{grad}

\DeclareMathOperator{\dom}{Dom} \DeclareMathOperator{\ran}{Ran}

\renewcommand{\div}{\mathrm{div}\,}

\DeclareMathOperator{\spec}{spec}

\DeclareMathSymbol{\newtimes}{\mathbin}{SY}{'264}

\newcommand{\rlad}{\mathrm{Re}^*}

\newcommand{\R}{\mathbb{R}}

\newcommand{\C}{\mathbb{C}}

\newcommand{\EE}{\mathsf{E}}

\newcommand{\fS}{\mathfrak{S}}

\newcommand{\fK}{\mathfrak{K}}

\newcommand{\fa}{\mathfrak{a}}
\newcommand{\fb}{\mathfrak{b}}
\newcommand{\fd}{\mathfrak{d}}

\newcommand{\ft}{\mathfrak{t}}
\newcommand{\fs}{\mathfrak{s}}
\newcommand{\fv}{\mathfrak{v}}

\newcommand{\cB}{{\mathcal B}}

\newcommand{\cD}{{\mathcal D}}

\newcommand{\cG}{{\mathcal G}}
\newcommand{\cH}{{\mathcal H}}

\newcommand{\cK}{{\mathcal K}}
\newcommand{\cL}{{\mathcal L}}

\newcommand{\sE}{{\mathsf E}}

\newcommand{\vv}{{v_{\ast}}}

{\bf}{\it}
{\bf}{\it}
{\bf}{\it}
{\bf}{\it}
{\bf}{\it}
{\bf}{\it}

{\bf}{\it}

\newtheorem{theorem}{Theorem}[section]{\bf}{\it}
\newtheorem{proposition}[theorem]{Proposition}{\bf}{\it}
{\bf}{\it}
\newtheorem{example}[theorem]{Example}{\it}{\rm}
\newtheorem{lemma}[theorem]{Lemma}{\bf}{\it}
\newtheorem{remark}[theorem]{Remark}{\it}{\rm}
\newtheorem{definition}[theorem]{Definition}{\bf}{\it}
{\bf}{\it}
{\bf}{\it}
{\bf}{\it}
\newtheorem{hypothesis}[theorem]{Hypothesis}{\bf}{\it}

\title[Diagonalization of indefinite saddle point forms]{Diagonalization of indefinite saddle point forms}

\author[L. Grubi\v{s}i\'c]{Luka Grubi\v{s}i\'c}
\address{L.~Grubi\v{s}i\'c,
Department of Mathematics, University of Zagreb, Bijeni\v{c}ka 30,
10000 Zagreb, Croatia}
\email{luka.grubisic@math.hr}

\author[V. Kostrykin]{Vadim Kostrykin}
\address{V.~Kostrykin, FB 08 - Institut f\"{u}r Mathematik,
Johannes Gutenberg-Universit\"{a}t Mainz,
Staudinger Weg 9,
D-55099 Mainz,
Germany}
\email{kostrykin@mathematik.uni-mainz.de}

\author[K. A. Makarov]{Konstantin A.~Makarov}
\address{K.~A.~Makarov, Department of Mathematics, University of
Missouri, Co\-lum\-bia, MO 65211, USA}
\email{makarovk@missouri.edu}

\author[S.~Schmitz]{Stephan Schmitz}
\address{S.~Schmitz, Department of Mathematics, University of
Missouri, Co\-lum\-bia, MO 65211, USA}
\email{schmitzst@missouri.edu}

\author[K. Veseli\'c]{Kre\v{s}imir Veseli\'c}
\address{K.~Veseli\'c,
Fakult\"{a}t f\"{u}r Mathematik und Informatik, Fernuniversit\"{a}t Hagen, Postfach 940,
D-58084 Hagen, Germany} \email{kresimir.veselic@fernuni-hagen.de}



\dedicatory{To the memory of Boris Sergeevich Pavlov}

\copyrightinfo{2017}{L.~Grubi\v{s}i\'c, V.~Kostrykin, K.~A.~Makarov, S.~Schmitz,
K.~Veseli\'c}
\begin{document}

\begin{abstract}
 We  obtain sufficient conditions that ensure block diagonalization   (by a  direct rotation) of sign-indefinite symmetric sesquilinear forms as well as the  associated operators that are semi-bounded neither from  below  nor from above. In the semi-bounded case, we refine the obtained results and, as an example,  
revisit the block Stokes Operator from fluid dynamics.\end{abstract}

\maketitle

\section{Introduction}

Diagonalizing a quadratic form, which is a classic problem of linear algebra and operator theory, is closely related to
the search for invariant subspaces for  the  (bounded)  operator associated with the form.
In the Hilbert space setting, 
a particular case where an invariant subspace can be represented as the graph of a bounded operator acting from a given subspace of the  Hilbert space to its orthogonal complement, is of special interest.
This situation is quite common  while  studying block operator matrices,  where  an  orthogonal decomposition of the  Hilbert space 
is available by default. In particular, solving the corresponding invariant graph-subspace problem  for bounded  self-adjoint block operator matrices 
automatically yields  a block diagonalization   of the matrix  by a unitary transformation.  
It is important to  note
 that  solving the problem  is completely nontrivial even in the bounded case: a self-adjoint operator matrix may have no invariant graph subspace (with respect to a given  orthogonal  decomposition) and, therefore, may not be  block diagonalized in this sense, see,  e.g., \cite[Lemma 4.2]{KMM03b}. 

To describe  the block diagonalization procedure in the self-adjoint bounded case in more detail, assume that the Hilbert space \(\cH\) splits into a direct sum of its subspaces,  \(\cH=\cH_+\oplus\cH_-\), and suppose  that  $ B  $  is  a $2\times 2$ self-adjoint  block operator matrix with respect to this decomposition.  In the framework of off-diagonal perturbation theory, we also assume that 
$ B =A+V $, with $ A$  and \(V\)  the diagonal and  off-diagonal parts of  $B$, respectively.  

 We briefly recall that the search for an invariant   subspace of $B$ that can be represented as the graph of  a bounded (angular) operator $X$ acting from $\cH_+$ to $\cH_-$
 is known to  be equivalent to finding   the skew-self-adjoint ``roots"
  \begin{equation*}
  Y := \begin{pmatrix} 0 & -X^\ast\\ X & 0 \end{pmatrix}
 \end{equation*}
 of the (algebraic) Riccati equation  (see, e.g., \cite{AMM,MSS})
 \begin{equation*}
 AY - YA - YVY + V = 0.
\end{equation*}

Given such a solution \(Y\), one observes that  the Riccati equation can be rewritten as the following operator equalities
$$(A+V)(I_\cH+Y) = (I_\cH+Y)(A+VY)
\;\text{ and }\;
 (I_\cH-Y)(A+V) = (A-YV)(I_\cH-Y),$$
with  $A+VY=(A-YV)^\ast$ block diagonal operators. 

In turn, those operator equalities ensure a block diagonalization of $B$ by the similarity transformation $I\pm Y$,  and,  as the next step, 
by the direct rotation $U$ from the subspace \(\cH_+\) to the invariant  graph subspace \(\cG_+=\mathrm{Graph}(\cH_+,X):=\{x+Xx\;|\;x\in \cH_+\}\)
 (see \cite{Davis,Davis:Kahan} for the concept of a direct rotation).
 Apparently, the direct rotation is given by the unitary operator from the polar decomposition 
\begin{equation*}
(I+Y)=U|I+Y|.\end{equation*} 

Solving the Riccati equation, the main step of the diagonalization procedure described above, attracted a lot of attention from several groups of researchers.

Different ideas and methods have been used to solve the Riccati equation under various assumptions on the (unbounded) operator \(B\).
 For an extensive list of references we refer to \cite{AMM} and \cite{T}
 (for matrix polynomial and Riccati equations in finite dimension  see   \cite{Egorov,FS, G2004,GLR,GLR2,LR, Plachenov}).
For  more recent results, in particular on Dirac operators with Coulomb potential, 
dichtonomous Hamiltonians, and bisectorial operators, we refer to \cite{Cue12,TW, JWZ, WW}.

The most comprehensive results regarding the solvability of the Riccati equation can be obtained under the hypothesis that the spectra of the diagonal part of the operator $B$ restricted to its reducing subspaces $\cH_\pm$ are subordinated. For instance, in the presence of a  gap separating the spectrum, the Davis-Kahan $\tan 2 \Theta$-Theorem \cite{Davis:Kahan}  can be used to  ensure
 the existence of contractive solutions to the corresponding Riccati equation. In this case,  efficient norm  bounds for the angular operator  can be obtained.
 The case where there is no spectral gap but the spectra of the diagonal entries have only one-point intersection $\lambda$   has also been treated, 
 see, e.g.,  \cite{ALT01, KMM:2},  \cite[Theorem 2.13]{SchPaper},  \cite[Proposition 2.7.13]{T}. Also, see  the recent work  \cite{MSS}, where, in particular, 
the decisive role  of  establishing  the kernel splitting  property 
\[\Ker(B-\lambda )=(\Ker(B-\lambda )\cap\cH_+)\oplus (\Ker(B-\lambda)\cap\cH_-)\] 
in the diagonalization process is discussed, cf. \cite[Section 2.7]{T}.

\vspace{12pt}

In the present paper, we extend the diagonalization scheme to the case of indefinite saddle point forms.
Recall that  a symmetric saddle point form $\fb$ with respect to the decomposition $\cH=\cH_+\oplus \cH_-$ is a form sum $\fb=\fa+\fv$, where   the diagonal part of the form \(\fa\) splits into the difference of two non-negative closed forms in the spaces \(\cH_+\) and \(\cH_-\), respectively,
and the off-diagonal part $\fv$ is a symmetric  form-bounded perturbation of $\fa$.

 First,  we treat the case 
where  the domain   of the form \(\dom[\fb]\)  and the form domain   \(\dom(|B|^{1/2})\) of the associated operator $B$ defined via the First Representation Theorem for  saddle point forms coincide. 
 Putting it differently, we assume that   the corresponding  Kato square root problem has an affirmative answer.
In this case, we show that on the one hand the semi-definite subspaces 
\begin{equation}\label{sese}
\cL_\pm=\ran \left  ( \EE_B(\mathbb{R}_\pm\setminus \{0\})\right )\oplus \big(\Ker (B)\cap \cH_\pm\big)
 \end{equation}
 reduce  both  the operator $B$ and  the form $\fb$.
  On the other hand, the semi-positive subspace $\cL_+$ is a 
 graph subspace  with respect to the decomposition $\cH=\cH_+\oplus\cH_-$,  see Theorem  \ref{vadi1}.
  Under some additional regularity assumptions, we  block diagonalize both the form and the associated operator by  the direct rotation from the subspace   \(\cH_+\) to the  subspace \(\cL_+\).

More generally, 
 we introduce the concept 
 of a
 block form  Riccati equation associated with   a given saddle point form
and relate its solvability to the existence of   graph subspaces that reduce the form.
Based on these  considerations,  we block diagonalize the   form by a unitary transformation, provided that some regularity requirements are met as well, see Theorem \ref{vadi3}.
\vspace{12pt}
 
 As an application,   we revisit the spectral theory 
 for  the Dirichlet  Stokes block operator 
 (that   describes   stationary   motion of a viscous fluid  in a bounded domain $\Omega\subset \mathbb{R}^d$)
  (see \cite{Reynolds}, cf.\ \cite{FFMM}).
\begin{equation*}
\begin{pmatrix}-\nu{\bf\Delta}&v^*\grad\\ -v^*\div &0\end{pmatrix}
\end{equation*}
in the direct sum of Hilbert spaces 
 $\cH=\cH_+\oplus\cH_-=L^2(\Omega)^d\oplus L^2(\Omega).$ 

\vspace{12pt}

The paper is organized as follows.

In Section 2, we introduce the class of saddle point forms  and  recall the corresponding  Representation Theorems for the associated operators.  

In Section 3, we  discuss  reducing subspaces for saddle point forms that  are the graph of a bounded operator.

In  Section 4, we recall the concept of a direct rotation and define the class of regular graph decompositions.

In Section 5,
we block diagonalize the associated operator by a unitary transformation  provided that the domain stability condition holds and  that the graph decomposition 
$
\cH=\cL_+\oplus\cL_-$  into the sum of semi-definite subspaces $\cL_\pm$ given by \eqref{sese} is regular, see Theorem \ref{vadi2}.

In Section 6, we introduce the concept of a block form Riccati equation and provide sufficient conditions for the  block diagonalizability of a saddle point form by a unitary transformation, see Theorem \ref{vadi3}.

In Section 7, we  discuss  semi-bounded saddle point forms
and  illustrate our  approach on an example from fluid dynamics. 

We adopt the following notation. 
In the Hilbert space \(\cH\) we use the scalar product \(\langle \;\cdot\;,\;\cdot\;\rangle\) semi-linear  the first and linear in the second component.
Various auxiliary quadratic forms
will be denoted by $\ft$. We write $\ft[x]$ instead of
$\ft[x,x]$. $I_{\fK}$ denotes the identity operator on a Hilbert space
$\fK$, where we frequently omit the subscript. 
If \(\ft\) is a quadratic form and \(S\) is a bounded operator we define the sum
\(\ft+S\) as the form sum \(\ft+\langle\:\cdot\:,S\:\cdot\:\rangle\) on the natural domain \(\dom[\ft]\).
For operators \(S\) and Borel sets \(M\) the corresponding spectral projection is denoted by \(\EE_S(M)\).
Given an orthogonal decomposition $\fK_0\oplus\fK_1$ of the Hilbert
space $\fK$ and dense subsets $\cK_i\subseteq\fK_i$, $i=0,1$, by
$\cK_0\oplus\cK_1$ we denote a subset of $\fK$ formed by the vectors
$\begin{pmatrix} x_0 \\ x_1 \end{pmatrix}$ with $x_i\in\cK_i$, $i=0,1$.
For a self-adjoint operator \(T\) we introduce the  corresponding Sobolev space \(\cH_T^1\) as the set \(\dom(|T|^{1/2})\)
equipped with the norm $
\left\|f\right\|_{\cH_T^1}=\sqrt{\left\||T|^{1/2}f\right\|^2+\left\|f\right\|^2}. $  
 
\subsection*{Acknowledgements} K.~A.~M.~ and S.~S.~ are indebted to Fritz Gesztesy, Alexandr Plachenov,  and Albrecht Seelmann for useful discussions.
L.~G.~ was partially supported by the grant HRZZ-9345 of the Croatian Science Foundation.
K.~A.~M.~ is indebted to the Institute for Mathematics for its kind hospitality during his one month stay at the Johannes
Gutenberg-Universit\"at Mainz in the Summer of 2014. The work of K.~ A.~ M.~ has been supported in part by the Deutsche
Forschungsgemeinschaft, grant KO 2936/7-1.


\section{Saddle point forms}

To introduce the concept of a saddle point form in a Hilbert space $\cH$, we pick up a self-adjoint involution $J$ given by the operator matrix \cite{pap:1},
\begin{equation}\label{decJ}
 J=\begin{pmatrix}I_{\cH_+}&0\\
 0&-I_{\cH_-}
 \end{pmatrix}_{\cH_+\oplus \cH_-}\end{equation}
with respect to a given  decomposition of the Hilbert space \(\cH\) into the orthogonal sum of its closed subspaces  
\begin{equation}\label{deccc}
\cH=\cH_+\oplus \cH_-.
\end{equation}

A sesquilinear form \(\fa\) is called diagonal (with respect to the decomposition \eqref{deccc} if the domain  \(\dom[\fa]\) is $J$-invariant and the form $\fa$ ``commutes" with the involution $J$, 
 $$\fa[x,Jy]=\fa[Jx,y]\quad \text{for}\quad x,y\in \Dom[\fa],$$
 and the form \begin{equation}\label{asubJ}\fa_J[x, y]= \fa[x, Jy]\quad \text{on}\quad \dom[\fa_J]=\dom[\fa]\end{equation} is a closed  non-negative form.
In particular, the form \(\fa\) splits into the difference of closed non-negative forms \(\fa=\fa_+\oplus (-\fa_-)\) with respect to the decomposition \(\cH=\cH_+\oplus \cH_-\).  

Correspondingly, a sesquilinear form \(\fv\) is called off-diagonal if  the ``anti-commutation relation" 
$$\fv[x,Jy]=-\fv[Jx,y]\quad  \text{for}\quad   x,y\in \Dom[\fv]$$
holds.

  We say that a form  $\fb$  is a saddle point form with respect to the decomposition \eqref{deccc}
if it admits the representation
$$\fb[x,y]=\fa[x,y]+\fv[x,y],\quad x,y\in \Dom[\fb]=\Dom[\fa],$$
where $\fa$ is a diagonal form, 
  $\fv$  is a symmetric  off-diagonal form
and relatively bounded with respect to \(\fa_J\),
\begin{equation*}
	|\fv[x]|\leq \beta(\fa_J[x]+\|x\|^2),\quad x\in \Dom[\fv],
\end{equation*}
for some $\beta\ge 0$.
 
 We start by recalling the First and Second Representation Theorem adapted here to the case of saddle point forms (see \cite[Theorem 2.7]{SchPaper}, \cite{pap:1}, \cite{pap:2}, see also \cite{McIntosh:1}).

\begin{theorem}[The First Representation Theorem]

\label{1repoffdiag}

Let \(\fb\) be  a saddle point form with respect to the decomposition  \(\cH=\cH_+\oplus \cH_-\).

Then there exists a unique self-adjoint operator  \(B\) such that  $$\Dom(B)\subseteq \Dom[\fb]$$  and 
\begin{equation*}
	\fb[x,y]=\langle x,By\rangle \quad\text{ for all }\quad x\in \Dom[\fb] \quad \text{ and }\quad y\in \Dom(B).
	\end{equation*}

\end{theorem}

We say that the operator $B$ associated with the saddle point form $\fb$  via Theorem  \ref{1repoffdiag}
satisfies the {\it domain stability condition} if the Kato square root problem has an affirmative answer. That is, 
\begin{equation}\label{domain stability}
\dom[\fb]=\dom(|B|^{1/2}).
\end{equation}

We note that the domain stability condition may fail to hold for form-bounded  but not necessarily off-diagonal perturbations of a diagonal form,  see \cite[Example 2.11]{pap:1}  and \cite{Fleige:3} for counterexamples.

The corresponding Second Representation Theorem can be stated as follows.  

\begin{theorem}[The Second Representation Theorem] 
\label{second}	Let \(\fb\) be  a saddle point form with respect to the decomposition  \(\cH=\cH_+\oplus \cH_-\) and 
 $B$ the associated operator.
	 
	 If the domain stability condition    \eqref{domain stability} holds, 
 then the operator \(B\)  represents this form in the sense that
\begin{equation*}\fb[x,y]=\langle |B|^{1/2}x,\sign(B)|B|^{1/2}y\rangle \quad \text{for all } x,y  \in \dom[\fb]=\dom(|B|^{1/2}).
\end{equation*}
\end{theorem}

\begin{remark}Let $\fa$ be a diagonal form and $A=JA_J$, where $A_J$ is a self-adjoint operator associated with the closed non-negative form $\fa_J$ in \eqref{asubJ}.
Clearly, the operator $A$ is associated with the form $\fa$ and  the form $\fa$ is represented by $A$ as well.
Notice that \(\fa_J\) is associated in the standard sense with the self-adjoint operator $|A|$.
\end{remark}

Next, we present  an  example of  a saddle point form   ``generated" by  an operator.

\begin{example}\label{illdef}
Given the decomposition 
\eqref{deccc},
suppose that \(A_\pm\ge 0\) are self-adjoint operators in \(\cH_\pm\).  Also suppose 
 that  $$W\colon\Dom(W)\subseteq\cH_+\to\cH_-$$  is a  densely defined closable
 linear operator such that 
  \begin{equation}\label{dommincl}
 \dom(A_+^{1/2})\subseteq\dom(W).
 \end{equation}

Let $\fa$  be the diagonal saddle point form associated with the diagonal operator 
  \begin{equation*}
  A=\begin{pmatrix} A_+ &0\\0 & -A_-\end{pmatrix}.
  \end{equation*}

On $$\dom [\fb]=\dom[\fa]=\dom(|A|^{1/2})$$  consider the form sum 
 \begin{equation}\label{a+b}
  \fb=\fa+\fv ,
 \end{equation}
  where the off-diagonal symmetric perturbation is given by
\begin{equation*}
\fv[x,y]=\langle Wx_+, y_-\rangle+\langle x_-,  Wy_+\rangle,
\end{equation*} 
$$
x=x_+\oplus x_-, \quad y=y_+\oplus y_-, \quad x_\pm, y_\pm \in \dom(|A_\pm|^{1/2})\subseteq \cH_\pm.
$$
\end{example}

\begin{lemma}\label{genex2}
 The form $\fb$ defined by \eqref{a+b} in Example \ref{illdef}  is a saddle point form.
Moreover, the off-diagonal part $\fv$ of the form $\fb$  is infinitesimally form-bounded with respect to the non-negative closed form $\fa_J$
given by $$\fa_J[x,y]=\langle |A|^{1/2}x, |A|^{1/2}y\rangle $$
 on $\Dom[\fa_J]=\Dom(|A|^{1/2})$. 

\begin{proof}
{}From \eqref{dommincl}
 it follows  that  the operator $W$ is  $A_+^{1/2}$-bounded  
(see \cite[Remark IV.1.5]{Kato}) and therefore
$$
\|W x_+\|\le a\|x_+\|+b\|A_+^{1/2}x_+\|, \quad x_+\in \Dom (A_+^{1/2}),
$$
for some  constants  $a$ and $b$. This shows the off-diagonal part \(\fv\) of the form \(\fb\) is relatively bounded with respect to the diagonal form  \(\fa_J\) and hence $\fb$ is a saddle point form.

 The last assertion follows from the series of inequalities
\begin{align*}
|\fv[x]|&\le 2|\langle W x_+, x_-\rangle|\le 2 \|Wx_+\|\cdot\|x_-\|\le 2\big(a\|x_+\|+b\|A_+^{1/2}x_+\|\big)\|x_-\| 
\\
&=2a\|x_+\|\cdot\|x_-\|+2b\sqrt{\fa_+[x_+]}\|x_-\|\nonumber\\
&\leq (a^2+1)\|x\|^2+ \varepsilon b^2\fa_+[x_+]+\frac{\|x\|^2}{\varepsilon} ,\nonumber
\end{align*}
$$
x=x_+\oplus x_-, \; x_\pm\in\dom[\fa]\cap \cH_\pm
$$
valid for all \(\varepsilon >0\).
\end{proof}
\end{lemma}

\begin{remark}\label{rem:illdefined}
The operator $B$ associated with the saddle point form $\fb$  from Example \ref{illdef} can be considered a self-adjoint realization of the ``ill-defined`` Hermitian  operator matrix 
  \begin{equation}
	\label{Bdot}
\dot B=\begin{pmatrix} A_+ & W^*\\ W & -A_-\end{pmatrix}. \end{equation}
 
Note that in this case we do not impose any condition on \(\dom(A_-)\cap \dom(W^\ast)\), so that the ``initial" operator \(\dot B\) is not necessarily densely defined
on its  natural domain \(\dom(\dot B)=\dom(A_+)\oplus\left(\dom(A_-)\cap \dom(W^\ast)\right)\).
 In particular,  we neither require that \(\Dom(A_-)\supseteq \Dom(W^*)\), cf.~ \cite{Atkinson}, nor that \(\dot{B}\) is essentially self-adjoint, cf.~\cite[Theorem 2.8.1]{T}. 
\end{remark}

We close this section  by the observation  that semi-bounded saddle point forms are automatically closed is the standard sense.

Recall that a  linear set \(\cD\subseteq\cH\) is called a core for the semi-bounded from below form \(\fb\geq c I_\cH\) if \(\cD\subseteq\dom[\fb]\) is dense in \(\dom[\fb]\) with respect to the norm \(||f||_\fb=\sqrt{\fb[f]+(1-c)||f||^2}\), see, e.g., \cite[Section VIII.6]{RS1}.
	
\begin{lemma}\label{lem:semibounded}
Suppose that $\fb$ is a  semi-bounded saddle point form   with respect to the decomposition  \(\cH=\cH_+\oplus \cH_-\). Then $\fb$ is closed in the standard sense. In particular,
the domain stability condition \eqref{domain stability} automatically holds.

Moreover, if   \(\cD\) is  a core for the diagonal part \(\fa\) of the form $\fb$,  then \(\cD\) is also a core for \(\fb\).

\begin{proof} 
Assume for definiteness that  \(\fb\) is semibounded from below. 
Let  $\fa$ and $\fv$   be  the diagonal and off-diagonal  parts of the form $\fb$, respectively. 

Since the off-diagonal part  \(\fv\) is relatively bounded with respect to \(\fa_J\),  that is,   \[|\fv[x]|\leq \beta(\fa_++\fa_-+I)[x]=\beta\langle (|A|+I)^{1/2}x,(|A|+I)^{1/2}x\rangle,\quad x\in \Dom[\fa],\]
for some $\beta<\infty$, applying  \cite[Lemma VI.3.1]{Kato}  shows that $\fv$ admits the representation 
 \begin{equation*}
 \fv[x,y]=\langle (|A|+I)^{1/2}x,R(|A|+I)^{1/2}y\rangle, \quad x,y\in\Dom[\fa],
 \end{equation*}
 with  a bounded operator \(R\).
  
Since \(\fv\) is off-diagonal,  the operator  \(R\) is off-diagonal as well, so that 
   $$
   JR=-RJ.
   $$
	
	Introducing the form 
   $$
   \widetilde \fb[x,y]= \fb[x,y]+\langle x,Jy\rangle, \quad x,y \in\Dom[\fa],
   $$ 
one observes  that 
\begin{equation*}
\widetilde
\fb[x,y]=\langle (|A|+I)^{1/2}x, (J+R)(|A|+I)^{1/2}y\rangle,\quad x,y \in\Dom[\fa].
\end{equation*}
Here we  used that
\begin{equation*}
\fa[x,y]+\langle x, Jy\rangle=\langle |A|^{1/2}x, J|A|^{1/2}y\rangle +\langle x, Jy\rangle=\langle (|A|+I)^{1/2}x, J(|A|+I)^{1/2}y\rangle
\end{equation*} for \(x,y \in\Dom[\fa]\). 

Since the spectrum  of $J$ consists of no more than two points $\pm1$ and the operator $R$ is off-diagonal,  the interval $(-1,1)$ belongs to the resolvent set of the bounded operator $J+R$. 
In particular, $J+R$ has a bounded inverse, see \cite[Remark 2.8]{KMM:2}. Since  \(|A|+I\) is strictly positive, applying  the First Representation Theorem \cite[Theorem 2.3]{pap:1}  shows  that the self-adjoint operator $\widetilde B
=(|A|+I)^{1/2} (J+R)(|A|+I)^{1/2}$  is associated with the semi-bounded form 
$\widetilde \fb$ and is semi-bounded as well.
Taking into account the one-to-one
 correspondence between closed semi-bounded forms and semi-bounded self-adjoint operators  proves that the form \(\widetilde{\fb}\) is closed, so is $\fb$ as a bounded perturbation of a closed form. 
 
To show that any core for the diagonal part \(\fa\) is also a core for \(\fb\), we remark first that  since \(\fb\) is  semi-bounded from below, the diagonal part $\fa$  of the form $\fb$ is semi-bounded from below as well.
Indeed, otherwise, the form \(\fa_-\) is not bounded and therefore there is a sequence \(x_n\in \dom[\fa_-], \left\|x_n\right\|=1\), such that 
\(\fa_-[x_n]\to \infty\). In this case, $$\fb[0\oplus x_n]=-\fa_-[x_n]\to -\infty, $$ which  contradicts the assumption that $\fb$ is a semi-bounded from below form.

Now, since  $\fb$ is  closed, by \cite[Theorem VI.2.23]{Kato}, the domain stability condition \eqref{domain stability} holds. This means that the Sobolev (Hilbert) spaces $\cH_A^1$ and $\cH_B^1$  associated with the operators \(A\) and \(B\) coincide.
Hence $\cD$ is dense in $\cH_A^1$ if and only if it is dense in $\cH_B^1$ (w.r.t. the  natural topology on the form domain). In other words,  \(\cD\) is a core for the form \(\fb\)  whenever it is  a core for the form \(\fa\).
\end{proof}
\end{lemma}

\section{Reducing subspaces} 
Recall that  a closed subspace 
$\fK$ reduces a self-adjoint operator $T$ if $$QT\subseteq TQ,$$
where  $Q$ is 
the orthogonal projection in $\cH$ onto  $\fK$  (see \cite[Section V.3.9]{Kato}).

This notion of a reducing subspace \(\fK\) means that \(\fK\) and its orthogonal complement \(\fK^\bot\) are invariant for \(T\) and  the domain of the operator \(T\) splits as \[\dom(T)=\big(\dom(T)\cap \fK\big)\oplus\big(\dom(T)\cap \fK^\bot\big).\]
  
Next, we introduce the corresponding notion for sesquilinear forms.

\begin{definition}\label{def:6:1} We say that a closed subspace $\fK$  of a Hilbert space \(\cH\) 
 reduces a symmetric
densely defined quadratic form $\ft$ with domain $\Dom[\ft]\subseteq\cH$ if
\begin{itemize}
\item[(i)]{$Q\left(\Dom[\ft]\right)\subseteq \Dom[\ft]$}
\item[] and 
\item[(ii)]{$\ft[Qu, v]=\ft[u,Qv]$  for all $u,v\in\Dom[\ft]$,}
\end{itemize}
where  $Q$ is the orthogonal projection
onto $\fK$.
\end{definition}

 A short computation shows that a closed subspace $\fK$ reduces a symmetric
densely defined quadratic form $\ft$ if
and only if 
\begin{equation}\label{rm:2:2}Q\left(\Dom[\ft]\right)\subseteq\Dom[\ft]\quad\text{and}\quad
\ft[Q^\perp u,Q
v]=0\;\text{ for all }u,v\in\Dom[\ft].
\end{equation}
In particular, $\fK$ reduces the form $\fb$ if and only if the orthogonal complement $\fK^\perp$ does.

Taking this into account,  along with  saying that a closed subspace $\fK$ reduces a form, we  also occasionally say that the orthogonal decomposition $\cH=\fK\oplus \fK^\perp$ reduces the form.

The following lemma shows that under the domain stability condition, the concepts of reducibility for the form and the associated (representing) self-adjoint operator coincide.

\begin{lemma}\label{redred} Assume that $\fb$ is a  saddle point form with respect to the decomposition \(\cH=\cH_+\oplus \cH_-\) and \(B\) the associated operator. 
Suppose that the domain stability condition  \eqref{domain stability} holds. 

Then a closed subspace  $\fK$ reduces the form \(\fb\) if and only if $\fK$ reduces the  operator \(B\).
\begin{proof}
Assume that $\fK$  reduces the form \(\fb\). Denote by \(Q\) the orthogonal projector onto \(\fK\).
In this case,  \[\dom[\fb]=\dom[\fa]=\dom(|B|^{1/2})\]
and
\[Q\left(\dom(|B|^{1/2})\right)\subseteq \dom(|B|^{1/2}).\]
Moreover, 
$$
\fb[Qx, y]=\fb[x,Qy]\quad \text{  for all} \;  x,y\in\Dom[\fb].
$$ 
Since the form $\fb $ is represented by $B$, we have
\[\langle |B|^{1/2}Qx, \sign(B)|B|^{1/2}y\rangle =\langle |B|^{1/2}x, \sign(B)|B|^{1/2}Qy\rangle\quad\text{for all }x,y \in \dom(|B|^{1/2}).\]
In particular, 
\[\langle Qx,By\rangle=\langle Bx,Qy\rangle\quad\text{ for all }x,y\in \dom(B).\]
Since \(B\) is self-adjoint, this means that 
$Qy\in \Dom(B)$ and that 
\[ QBy=BQy \quad\text{ for all }\;x\in \dom(B),\]
which shows that 
 $\fK$  reduces the self-adjoint operator \(B\).

To prove the converse, suppose that  $\fK$  reduces the operator \(B\). By \cite[Satz 8.23]{Wei}, the decomposition also reduces  both operators \(|B|^{1/2}\) and \(\sign(B)\). Together with \[\dom[\fb]=\dom[\fa]=\dom(|B|^{1/2})\] this means that  
$$Q\left(\dom[\fb]\right)\subseteq\dom[\fb]$$ and that  \(Q\) commutes with \(\sign(B)\) and \(|B|^{1/2}\). Thus, 
\[\fb[Qu,v]= \langle |B|^{1/2}Qu,\sign(B)|B|^{1/2}v\rangle=\langle |B|^{1/2}u,\sign(B)|B|^{1/2}Qv\rangle=\fb[u,Qv],\]
which shows that $\fK$ reduces  the form \(\fb\).\qedhere
\end{proof}
\end{lemma}

The  theorem below generalizes of a series of results of \cite{AL95, ALT01, KMM:2, SchPaper}, cf. \cite[Section 2.7]{T}, and provides a canonical  example
of a semi-definite reducing subspace for a saddle-point form.

\begin{theorem}\label{vadi1}
 Let \(\fb\) be  a saddle point form with respect to the orthogonal decomposition $\cH=\cH_+\oplus \cH_-$ and $B$ the associated operator. 
 Assume that the form $\fb$ satisfies the domain stability condition \eqref{domain stability}. 

Then the subspace \(\Ker(B)\cap \cH_+\) reduces both the form $\fb$ and the operator \(B\). In particular, the kernel of \(B\) splits as \[\Ker(B)=\left(\Ker(B)\cap \cH_+\right)\oplus \left(\Ker(B)\cap \cH_-\right),\]
the semi-definite subspaces \begin{equation*}
\cL_\pm=\left (\ran \EE_B((\mathbb{R}_\pm\right )\setminus \{0\}))\oplus \big(\Ker (B)\cap \cH_\pm\big) 
  \end{equation*} 
  are complimentary, and the orthogonal decomposition
   \begin{equation*}
   \cH=\cL_+\oplus\cL_-
   \end{equation*}
   reduces both the form $\fb$ and the associated operator $B$. 

Moreover, the subspace $\cL_+$  is a graph of a linear contraction   $X\colon\cH_+\to\cH_-$.
 \begin{proof}
 Assume  temporarily that \(\Ker(B)=\{0\}\). Then the orthogonal decomposition \(\cH=\cL_+\oplus \cL_-\) is spectral. Therefore \(\cL_+\) and \(\cL_-\) reduce the operator \(B\) and, by the domain stability condition and Lemma \ref{redred}, the form \(\fb\) as well.

To complete the proof under the assumption \(\Ker(B)=\{0\}\),  we check that \(\cL_\pm\) are graph subspaces. Denote by $P$ the orthogonal projection onto $\cH_+$ and let \(Q=\sE_B(\mathbb{R}_+)\) be the spectral projection of \(B\) onto its positive subspace.
Introducing the sequence of self-adjoint operators
$$
 B_n=B+\frac1nJ,\quad J= \begin{pmatrix}I_{\cH_+}&0\\0&-I_{\cH_-}\end{pmatrix}, \quad n\in \mathbb{N}, 
$$
one observes that 
$$
\lim_{n\to \infty}B_n\varphi =B\varphi, \quad \varphi \in \Dom(B).
$$

By \cite[Theorem VIII.25]{RS1}, the sequence of  operators $B_n$ converges to $B$ in the strong resolvent sense, and therefore, by  \cite[Theorem VIII.24]{RS1},
 \begin{equation}\label{slimm}\slim_{n \to \infty}\sE_{B_n}(\mathbb{R}_+)=\sE_B(\mathbb{R}_+),
 \end{equation}
 since $0$ is not an eigenvalue of $B$.
 
 Taking into account that the operator $B_n$ is associated with the form $\fb_n$
 given by \[\fb_n[x,y]:=\fb[x,y]+\frac{1}{n}\langle x,Jy\rangle\]  and that  the interval \((-1/n,1/n)\) belongs to  its resolvent set, one applies the Tan\ \(2\Theta\)-Theorem \cite[Theorem 3.1]{pap:2}  to conclude that 
 \begin{equation}\label{eq:preLimit}
 \|Q-\EE_{B_n}(\mathbb{R}_+)\|<\frac{\sqrt{2}}{2}.
 \end{equation}

Since \eqref{slimm} holds, one also gets the weak limit 
\begin{equation}\label{eq:limit}
\wlim_{n \to \infty}(Q-\EE_{B_n}(\mathbb{R}_+))=Q-\EE_{B}(\mathbb{R}_+).
\end{equation}
 Using   the principle of uniform boundedness, see \cite[Equation (3.2),  Chapter III]{Kato},
one  obtains from  \eqref{eq:preLimit} and \eqref{eq:limit} the bound
$$
\|Q-\EE_{B}(\mathbb{R}_+)\|\le \liminf_{n\to \infty}\|Q-\EE_{B_n}(\mathbb{R}_+)\|\le \frac{\sqrt{2}}{2}.
$$
Hence,  \(\cL_+\) is the graph subspace $\mathrm{Graph}(\cH_+, X)$ with  $X$ a contraction, see \cite[Corollary 3.4]{KMM:1}. The orthogonal complement \(\cL_-\) is then the  graph subspace  $\mathrm{Graph}(\cH_-, -X^*)$.  

\vspace{6pt}

We now treat the general case (of a non-trivial kernel). 

\vspace{6pt}
First, we check that the semi-positive subspaces \(\cL_\pm\) reduce the operator \(B\), and thus also the form \(\fb\).

It is clear that both \(\cL_\pm\) are invariant for \(B\).
It is also clear that  the subspaces \(\cL_\pm\) are complimentary if and only if the kernel splits  as
\begin{equation}\label{eq:splittheKernel}\Ker(B)=(\Ker(B)\cap \cH_+)\oplus(\Ker(B)\cap \cH_-).\end{equation}

 To prove \eqref{eq:splittheKernel}, recall (see \cite[Theorem 2.13]{SchPaper}) that the kernel of \(B\) can be represented as 
\begin{equation}\label{eq:kernel}\Ker(B)=(\Ker(A_+)\cap \cK_+)\oplus(\Ker(A_-)\cap \cK_-),\end{equation}
where $A_\pm$ are self-adjoint non-negative operators associated with the forms
$\fa_\pm$ and the subspaces \(\cK_+\) and \(\cK_-\) are given by 
\begin{equation*}
\cK_\pm:=\big\{x_\pm\in \dom[\fa_\pm]\;| \; \fv[x_+, x_-]=0 \quad \text{for all }x_\mp\in \dom[\fa_\mp]\big\}\subseteq\cH_\pm.\end{equation*}
Hence 
 \(\Ker(B)\cap \cH_\pm=\Ker(A_\pm)\cap\cK_\pm\)  and 
 \eqref{eq:splittheKernel} follows.

Next, in view of  \eqref{eq:splittheKernel}, since \(\cH\) naturally splits as
\[\cH=\Ran \EE_B(\mathbb{R}_+)\oplus\Ker(B)\oplus \Ran \EE_B(\mathbb{R}_-),\] 
one gets
\begin{align*}
\Dom(B)&=\big(\Dom(B)\cap \Ran \EE_B(\mathbb{R}_+)\big)\oplus (\Ker (B) \cap \cH_+)
\\&\quad\oplus(\Ker (B) \cap \cH_-)
\oplus \big(\Dom(B)\cap \Ran \EE_B(\mathbb{R}_+)\big).
\end{align*}
This  representation shows
that the domain $\Dom(B)$ splits as 
\begin{equation}\label{spsp}
\dom(B)=\big(\dom(B)\cap \cL_+\big)\oplus \big(\dom(B)\cap \cL_-\big).
\end{equation} 

Summing up, we have shown that $\cL_\pm$ are $B$-invariant mutually orthogonal subspaces
such that \eqref{spsp} holds. That is, the subspaces $\cL_\pm $ reduce the operator $B$ and therefore the form $\fb$.

To complete the proof, we now need to check that \(\cL_+\) (and thus also \(\cL_-\)) is a graph subspace with a contractive angular operator.

By \cite[Corollary 3.4]{KMM:1}, it it sufficient to show that  \begin{equation}\label{newsubspaceestimate*}
\left\|Q-P\right\|\le\frac{\sqrt{2}}{2},\end{equation}
   where $Q$  and $P$ are the orthogonal projection onto $\cH_+$ and $\cL_+$, respectively. 
   
We will prove  \eqref{newsubspaceestimate*} by reducing the problem to the one where the kernel is trivial.

First we show that   \(\Ker(B)\) reduces the operator \(A\). 
Indeed, 
by \eqref{eq:kernel} we have $\Ker(B)\subseteq\Ker(A),$ so that $\Ker(B)$
 is invariant for $A$. Hence, $\Ker(B)^\bot$ is invariant for $A$ as well. 
 It remains to check  that $\Dom(A)$ splits as 
 \begin{equation}\label{splitt}
\Dom(A)=( \Dom(A)\cap \Ker(B))\oplus ( \Dom(A)\cap \Ker(B)^\bot).
 \end{equation}
 Indeed, since $\Ker(B)$ reduces $B$,  by \cite[Satz 8.23]{Wei}, the subspace $\Ker (B)$ also reduces  \(|B|^{1/2}\). By the required domain stability condition, this implies that \(\Ker(B)\) reduces \(|A|^{1/2}\) and, by \cite[Satz 8.23]{Wei} again, also \(|A|\). Thus \eqref{splitt} holds  by observing that $\Dom(A)=\Dom(|A|)$.

We now complete the proof that \(\cL_+=\mathrm{Graph}(\cH_+,X)\) is a graph subspace for a contraction \(X\).
 
 Taking into account that  the subspace $\widetilde{\cH}:=\Ker(B)^\perp$ reduces both $A$ and $B$, denote by
 $\widetilde{A}:=A|_{\widetilde{\cH}}$ and $\widetilde{B}:=(B)|_{\widetilde{\cH}}$ the corresponding parts of $A$ and $B$, respectively. In particular,
 $\widetilde{A}$ and $\widetilde{B}$ are self-adjoint operators  and $\Ker(\widetilde{B})=\{0\}$.

 In view of the kernel splitting \eqref{eq:kernel}, a simple reasoning shows that $\widetilde{\cH}$
 splits  as
 \[
  \widetilde{\cH}=\widetilde{\cH}_+\oplus\widetilde{\cH}_-\quad  \text{with}\quad 
  \widetilde{\cH}_+:=\cH_+\cap\widetilde{\cH} \quad\text{and}\quad \widetilde{\cH}_-:=\cH_-\cap\widetilde{\cH}\,,
 \]
 and that the operator $\widetilde{A}$ is represented as the diagonal block matrix
 \[
  \widetilde{A} = \begin{pmatrix} \widetilde{A}_+ & 0\\ 0 & -\widetilde{A}_- \end{pmatrix}_{\widetilde\cH_+\oplus\widetilde\cH_-}
 \]
with
 \[
  \sup \spec(-\widetilde{A}_-) \le 0 \le \inf \spec(\widetilde{A}_+) \,.
 \]
 In this case the corresponding sesquilinear symmetric form \(\widetilde\fa\) also splits into the difference of two non-negative forms.
 The restriction  \(\widetilde\fb=\fb|_{\widetilde{\cH}}\) is clearly seen to be a saddle point form associated with  
the self-adjoint operator \(\widetilde B\). 
 Since  \(\Ker(\widetilde B)=\{0\}\), by the above reasoning, we get the inequality 
 $$
 \|\widetilde Q-\EE_{\widetilde{B}}\bigl(\mathbb{R}_+\bigr)\|\le \frac{\sqrt{2}}{2},
 $$
 where $\widetilde Q$ is the orthogonal projection onto $\widetilde \cH_+$ and \(\EE_{\widetilde{B}}\bigl(\mathbb{R}_+\bigr)\) is the spectral projection of \(\widetilde{B}\) for the positive part. In particular, as in the previous case, 
 $\Ran\EE_{\widetilde{B}}\bigl(\mathbb{R}_+\bigr)= \mathrm{Graph}(\widetilde{\cH}_+,\widetilde{X})$
  is the  graph  of  a linear contraction
 $\widetilde{X}\colon\widetilde{\cH}_+\to\widetilde{\cH}_-$.
 
Denoting by $X$  the extension of the operator \(\widetilde{X}\) by zero on \(\Ker(B)\cap\cH_+\) and taking into account that by \eqref{eq:kernel}
$$
A|_{\Ker(B)\cap\cH_+}=B|_{\Ker(B)\cap\cH_+}=0, 
$$we obviously get that  \(\cL_+=\mathrm{Graph}(\cH_+, X)\). Observing that the extended operator $X$  is  also a contraction completes the proof.
\end{proof}
\end{theorem}

\section{Regular embeddings and  Direct Rotations}

 Given the orthogonal decomposition 
 $$
 \cH=\cH_+\oplus \cH_-,
 $$
suppose  that  Hilbert spaces $\dot\cH_\pm$ are continuously embedded in $\cH_\pm$, \begin{equation}\label{Hdot}\dot \cH_\pm\hookrightarrow \cH_\pm,\end{equation}  so that their direct sum   \(\dot\cH=\dot\cH_+\oplus\dot\cH_-\) is also continuously embedded in $\cH=\cH_+\oplus\cH_-$.

Suppose that a subspace  $\cG_+$ can be represented as a graph of a bounded operator \(X\) from $\cH_+$ to $\cH_-$ and let
\begin{equation}\label{bach}
\cH=\cG_+\oplus\cG_-
\end{equation}
be the corresponding decomposition with $\cG_-=(\cG_+)^\perp$, the graph of the bounded operator \(-X^\ast\colon \cH_-\to \cH_+\).

\begin{definition} \label{def:reg}  
We say that the graph decomposition \eqref{bach} is $\dot\cH$-regular (with respect to the embedding) if the linear sets
\begin{equation*}
\dot\cG_\pm=\cG_\pm\cap \dot\cH
\end{equation*}
naturally embedded in $\dot\cH$ are closed complimentary  graph subspaces in the Hilbert space  $\dot\cH$ with respect to the decomposition $\dot\cH=\dot\cH_+\oplus \dot\cH_-$.
\end{definition}

Denote by  $P$ and $Q$ the orthogonal projections onto the subspaces $\cH_+$ and $\cG_+$, respectively.  

Recall that as long as it is known that $\cG_+$ is a graph subspace,
there exists a unique unitary operator \(U\) on \(\cH\) that maps $\cH_+$ to $\cG_+$, such that 
$$
UP=QU,
$$
the diagonal entries of which (in its block matrix representation with respect to the decomposition $\cH=\cH_+\oplus\cH_-$) are non-negative operators, see \cite{Davis:Kahan}. 
In this case the  operator $U$ is called the direct rotation from the subspace $\cH_+=\ran(P)$ to the subspace $\cG_+=\ran(Q)$.

\begin{lemma}\label{Umapsfine}

Suppose that the graph decomposition  \eqref{bach} is \(\dot\cH\)-regular with respect to the embedding \eqref{Hdot}. 
Let $U$ and $\dot U$ be the direct rotation from $\cH_+$ to $\cG_+$  in the space \(\cH\) and from   $\dot\cH_+$  to $\dot\cG_+$ in \(\dot\cH\), respectively. 
Then
 $$\dot U=U|_{\dot\cH}\;.$$ 
\begin{proof}

Since the graph decomposition  \eqref{bach} is \(\dot\cH\)-regular, it follows that \(\cG_+\cap \dot\cH\) is the graph of a bounded operator $\dot X:\dot \cH_+\to \dot \cH_-$. Therefore,
 \(\cG_+\cap \dot\cH\) is the graph of $-\dot X^*$. Clearly, 
 $$
 \dot X=X|_{\dot \cH_+}\quad \text{and}\quad (-\dot X^*)=(-X^*)|_{\dot \cH_-}.
 $$
 In particular, 
 $$
 X^*X|_{\cH_+}=\dot X^*\dot X \quad \text{and}\quad  XX^*|_{\cH_-}=\dot X\dot X^*.
 $$
A classic Neumann series argument  shows that
\begin{equation}\label{prom}
  (tI+X^*X)^{-1}|_{\cH_+}= (tI+\dot X^*\dot X)^{-1}
\end{equation}for $|t|$ is large enough.
Taking into account the continuity of the embedding, one extends \eqref{prom} 
for all $t>0$  by analytic continuation. Next,  using the formula for the fractional power (see, e.g., \cite[Ch. V, eq. (3.53)]{Kato})
$$
T^{-1/2}=\frac{1}{\pi}\int_0^\infty t^{-1/2}(T+tI)^{-1}dt
$$
valid for any positive self-adjoint operator $T$  and taking     $T= (I+X^*X)|_{\cH_+}$ first  and then $T=I+\dot X^*\dot X$ in the Hilbert spaces  $\cH_+$ and $  \cH$, respectively,   from  \eqref{prom}
one deduces that 
\begin{equation}\label{aaa}
  (I+X^*X)^{-1/2}|_{\cH_+}= (I+\dot X^*\dot X)^{-1/2}
\end{equation}
Analogously,
 \begin{equation}\label{bbb}
  (I+XX^*)^{-1/2}|_{\cH_-}= (I+\dot X\dot X^*)^{-1/2}.
\end{equation}

Since  the direct rotation $U$
admits the representation 
\begin{equation*}
U=\begin{pmatrix} (I+X^*X)^{-1/2}& -X^*(I+XX^*)^{-1/2}\\
X(I+X^*X)^{-1/2}& (I+XX^*)^{-1/2}\end{pmatrix},
\end{equation*}
 cf.~ \cite{AMM,Davis:Kahan,KMM:1}, see also \cite[Proof of Proposition 3.3]{Albrecht}, and analogously 
 \[\dot U=\begin{pmatrix} (I+\dot X^* \dot X)^{-1/2}& -\dot X^*(I+\dot X\dot X^*)^{-1/2}\\
 \dot X(I+\dot X^* \dot X)^{-1/2}& (I+\dot X\dot X^*)^{-1/2}\end{pmatrix},\] 
the assertion follows from \eqref{aaa} and \eqref{bbb}.
\end{proof}
\end{lemma}

\begin{remark}\label{rem:regularity}  If $\cG_+$ is the graph of a bounded operator $X$ from $\cH_+$ to $\cH_-$, introduce 
\begin{equation*}
Y=\begin{pmatrix}
0&-X^*\\
X&0
\end{pmatrix}_{\cH_+\oplus\cH_-}.\end{equation*} Then the direct rotation \(U\) is just the unitary operator from 
the polar decomposition  of the operator $I+Y$, 
\begin{equation*}
(I+Y)=U|I+Y|.
\end{equation*}
 Observe that the \(\dot\cH\)-regularity of the decomposition \(\cH=\mathrm{Graph}(\cH_+,X)\oplus \mathrm{Graph}(\cH_-,-X^\ast\)) can equivalently be reformulated in purely algebraic terms that invoke mapping properties of the operators $I\pm Y$ only.
That is, the graph space decomposition \eqref{bach} 
is \(\dot\cH\)-regular if and only if the operators   \(I\pm Y\) are algebraic/ topologic automorphisms of \(\dot\cH\), see \cite[Lemma 3.1, Remark 3.2]{MSS}.

\end{remark}

\section{Block-Diagonalization of Associated Operators by a Direct Rotation}\label{sec:diagonalization}

One of the main results  of the current  paper is as follows.

\begin{theorem}\label{vadi2}
Let \(\fb\) be  a saddle point form with respect to the orthogonal decomposition $\cH=\cH_+\oplus \cH_-$ and $B$ the associated operator. 
 Assume that the form $\fb$ satisfies the domain stability condition \eqref{domain stability}. 

Suppose that  the decomposition 
\begin{equation}\label{lala}
\cH=\cL_+\oplus \cL_-
\end{equation}
referred to in Theorem   \ref{vadi1} is \(\cH_A^1\)-regular.
 
Then  the form \(\fb\) and the associated operator  \(B\) can be block diagonalized by the direct rotation \(U\) from the subspace $\cH_+$ to the reducing graph subspace 
$\cL_+$. That is,  
\begin{itemize}
\item[(i)] the form
 $$
\widehat \fb[f,g]=\fb[Uf,Ug], \quad f,g\in \Dom[\widehat \fb]=\Dom[\fb]$$
  is a diagonal saddle point form with respect to the decomposition \(\cH=\cH_+\oplus \cH_-\),
  $$\widehat \fb=\widehat \fb_+\oplus(-\widehat \fb_-),$$ with \(\widehat{\fb}_\pm=\pm\widehat{\fb}|_{\cH_\pm}\). In particular, the non-negative the forms \(\widehat{\fb}_\pm\) are closed;
\item[(ii)]  
  the associated operator $\widehat B$ can be represented  as  the diagonal operator matrix,
 \begin{equation*}
\widehat B=U^*BU=\begin{pmatrix}\widehat B_+&0\\0&-\widehat B_-
\end{pmatrix}_{\cH_+\oplus\cH_-};
\end{equation*} 

\item[(iii)]	the non-negative closed  the forms \(\widehat{\fb}_\pm\) are in one-to-one correspondence to the non-negative self-adjoint operators \(\widehat{B}_\pm\).
\end{itemize}	

If, in addition, the form $\fb$ is semi-bounded, then the hypotheses that  
 $\fb$ satisfies the domain stability condition  and that  the  decomposition 
\eqref{lala} is regular are redundant.

\begin{proof}

Note that in the Hilbert space \(\cH_A^1\) the form \(\fb\) can be represented by a bounded operator \(\cB\), such that
\begin{equation}\label{eq:fbxy}\fb[x,y]=\langle x,\cB y\rangle_{\cH_A^1}, \quad x,y \in \dom[\fb].\end{equation}

Let \(\dot U\) denote the direct rotation from \(\cH_+\cap \cH_A^1\) to \(\cL_+\cap \cH_A^1\) in the Sobolev space \(\cH_A^1\). 
By Lemma \ref{Umapsfine}  one has \(\dot U=U|_{\cH_A^1}\) and therefore 
 \[\fb[Ux,Uy]=\langle \dot Ux,\cB\dot Uy\rangle_{\cH_A^1}=\langle x,(\dot U)^{*} \cB\dot Uy\rangle_{\cH_A^1}.\] 
Since the decomposition \eqref{lala} reduces \(\fb\), it follows that \((\dot U)^{*} \cB\dot U\) is a diagonal operator matrix in the Sobolev space  \(\cH_A^1\) with respect to the decomposition \(\cH_A^1=\left(\cH_{+}\cap \cH_A^1\right)\oplus\left( \cH_{-}\cap \cH_A^1\right)\). The corresponding subspaces \(\cL_\pm\) are non-negative subspaces for the operator \(B\), so that  \[(\dot U)^{*} \cB\dot U=\begin{pmatrix}\cB_+&0\\0& -\cB_-\end{pmatrix}_{\left(\cH_{+}\cap \cH_A^1\right)\oplus\left( \cH_{-}\cap \cH_A^1\right)},\] where \(\cB_\pm\)  are non-negative  bounded operators in \(\cH_\pm\cap \cH_A^1\). Since  \begin{equation*}\label{proofI}\fb[Ux_\pm, Uy_\pm]=\pm\langle x_\pm,  \cB_\pm y_\pm\rangle_{\cH_A^1\cap \cH_\pm},\end{equation*} one observes that 
\(\fb[Ux_\pm, Uy_\pm]\), \(x_\pm,y_\pm\in \cH_\pm\cap \cH_A^1\) defines a sign-definite closed form  on \(\cH_\pm\cap \cH_A^1\). 
This proves (i).

On the other hand, \[\fb[Ux,Uy]=\langle x, U^*BUy\rangle_{\cH}, \quad x\in \dom[\fb],\; y\in U^{-1}(\dom(B)).\]  

In particular, one has \begin{equation}\label{proofII}\fb[Ux_\pm,Uy_\pm]=\langle x_\pm,\pm \widehat{B}_\pm y_\pm\rangle_{\cH}, \quad x_\pm\in \dom[\fa_\pm],y_\pm\in U^{-1}(\dom(B))\cap \cH_\pm,\end{equation} 
which shows (ii).

The assertion (iii) now follows from (i) and \eqref{proofII}.

Next, we prove the last assertion of the theorem. Denote by \(Q\) the orthogonal projection onto the subspace \(\cL_+\). Then \begin{equation}\label{P}Q=\begin{pmatrix} (I_{\cH_+}+X^\ast X)^{-1} & (I_{\cH_+}+X^\ast X)^{-1}X^\ast\\ 
X(I_{\cH_+}+X^\ast X)^{-1}& X(I_{\cH_+}+X^\ast X)^{-1}X^\ast\end{pmatrix}
\end{equation}
and
\begin{equation}\label{Pbot}Q^\bot =\begin{pmatrix} X^\ast(I_{\cH_-}+XX^\ast)^{-1}X & -X^\ast (I_{\cH_-}+XX^\ast)^{-1}\\ 
-(I_{\cH_-}+XX^\ast)^{-1}X& (I_{\cH_-}+X X^\ast)^{-1}\end{pmatrix}.
\end{equation}

 Note that since the subspace \(\cL_+\) reduces the form \(\fb\), both \
\(Q\) and \(Q^\bot\) map \(\dom[\fb]\) into itself. In particular, \((I_{\cH_+}+X^\ast X)^{-1}\) maps \(\cH_-\cap \dom[\fb]\) into itself.

Now, since the form \(\fb\) is semibounded from below, the operator \(I+XX^\ast\) is bijective on \(\cH_-\cap \dom[\fb]\).
Therefore, 
the operator
$$
(I-Y^2)^{-1}=
\begin{pmatrix}
(I_{\cH_+}+X^* X)^{-1}&0\\
0&(I_{\cH_-}+X X^\ast)^{-1}
\end{pmatrix}
$$
maps \(\dom[\fb]\) into itself.

Again, since \(I+XX^\ast\) is bijective on $\cH_-=\dom[\fb]\cap\cH_-$ and $Q^\perp$ maps \(\dom[\fb]\) into itself, it follows from \eqref{Pbot} that
\(X\) maps   into  $\dom[\fb]\cap\cH_+$ into 
$\dom[\fb]\cap\cH_-$
 and \(X^*\) maps $\dom[\fb]\cap\cH_-$ into $\dom[\fb]\cap\cH_+$.  Thus,   \(Y\) leaves the form domain $\dom[\fb]$  invariant and so do the operators \(I+Y\), \(I-Y\) and \(I-Y^2\). 
 
 Summing up, both
 \((I-Y^2)\)  and  \((I-Y^2)^{-1}\) map \(\dom[\fb]\) into itself. That is, the restriction   of the map  
 $$I-Y^2=(I-Y)(I+Y)$$ on \(\dom[\fb]\) is bijective on \(\dom[\fb]\). In particular $I+Y$ is bijective on \(\dom[\fb]\) and by Remark \ref{rem:regularity} the decomposition $\cH=\cL_+\oplus\cL_-$ is \(\cH_A^1\)-regular, which completes the proof.

\end{proof} 
\end{theorem}

\section{The Riccati Equation}

The existence of a reducing  graph subspace for a saddle point form, as, for instance,  in Theorem \ref{vadi2}, is closely related to the solvability 
of the associated block form Riccati equation. 

\begin{hypothesis}\label{tech} Suppose that 
\(\fb\) is  a saddle point form with respect to  the decomposition \(\cH=\cH_+\oplus \cH_-\). Assume that a subspace  $\cG_+$ is the graph of a bounded operator $X:\cH_+\to \cH_-$ and that 
$Y$ is  the skew-symmetric off-diagonal operator $Y$
\begin{equation*}
Y=\begin{pmatrix}
0&-X^*\\
X&0
\end{pmatrix}_{\cH_+\oplus\cH_-}
\end{equation*}

\end{hypothesis}

\begin{theorem}\label{regularSaddlepointForm*}   Assume Hypothesis \ref{tech}.
 Assume, in addition,  that the orthogonal decomposition \(\cH=\cG_+\oplus\cG_-\)  is  \(\cH_A^1\)-regular. 

Then the  decomposition \(\cH=\cG_+\oplus\cG_-\) reduces the form 
 \(\fb\) if and only if the skew-symmetric off-diagonal operator $Y$
 is a solution to the block form Riccati equation
 \begin{equation}\label{RicBlock}\fa[f,Yg]+\fa[Yf,g]+\fv[Y f,Yg]+\fv[f,g]=0 , \quad f,g\in \dom[\fa],
\end{equation}
\begin{equation*}
\Ran(Y|_{\dom [\fa]})\subseteq \dom[\fa].
\end{equation*}

\begin{proof}   
The proof of this theorem is a direct combination of the following two lemmas.
\end{proof}
\end{theorem}

\begin{lemma}\label{redric} Assume Hypothesis \ref{tech} and suppose that \(\cH=\cG_+\oplus\cG_-\) reduces \(\fb\).

If  the Sobolev space $\cH_A^1$ is $Y$-invariant, then
  $Y$ is a solution to the block form Riccati equation \eqref{RicBlock}.
\begin{proof}  

Assume that the decomposition reduces  \(\fb\). Since \(\dom[\fb]=\dom[\fa]=\cH_A^1\) (as a set),
the $Y$-invariance  of the Sobolev space \(\cH_A^1\) implies that \(X\) and \(X^\ast\) map \(\dom[\fa_+]=\dom(A_+^{1/2})\) into \(\dom[\fa_-]=\dom(A_-^{1/2})\), and vice versa,  respectively. 
Denote by $Q$  the orthogonal projection onto  $\cG(\cH_+,X)$.
By \eqref{rm:2:2}, we have 
\begin{equation}\label{restat}
0=\fb[Q^\perp(-X^\ast y\oplus y),Q(x\oplus Xx)]=
\fb[-X^\ast y\oplus y,x\oplus Xx],
\end{equation}
$$x\in \dom[\fa_+],\;y\in \dom[\fa_-].
$$

Taking into account that 
 $\fb=\fa+\fv$, where  \(\fa\) and \(\fv\) are the diagonal and off-diagonal parts, respectively,  and  that $\fa=\fa_+\oplus(-\fa_-)$, the  equality \eqref{restat} shows that \(X\) is a solution to the Riccati  equation 
\begin{equation}\label{ric*}
\fa_+[-X^\ast y,x]-\fa_-[y,Xx]+\fv[-X^\ast y, Xx]+\fv[ y,x]=0,
\end{equation}
$$x\in \dom[\fa_+],\quad y\in \dom[\fa_-].
$$

Set \[f=x_+\oplus x_-,\;g=y_+\oplus y_-,\quad x_\pm,\;y_\pm\in \dom[\fa_\pm],\] 
combine the  Riccati equation \eqref{ric*} with  \(x=y_+,\; y=x_-\) plugged in, and  the complex conjugate 
of \eqref{ric*}  with \(x=x_+,\; y=y_-\) plugged in, to get
\begin{align*}
\fa[f&,Yg]+\fa[Yf,g]+\fv[Y f,Yg]+\fv[f,g]\\
&=\fa\left [\begin{pmatrix}x_+\\ x_-\end{pmatrix},\begin{pmatrix} 0&-X^*\\X&0\end{pmatrix}
\begin{pmatrix}y_+\\ y_-\end{pmatrix}\right ]+
\fa\left [\begin{pmatrix} 0&-X^*\\X&0\end{pmatrix}\begin{pmatrix}x_+\\ x_-\end{pmatrix},\begin{pmatrix}y_+\\ y_-\end{pmatrix}\right]
\\& \quad +
\fv\left [\begin{pmatrix} 0&-X^*\\X&0\end{pmatrix} \begin{pmatrix}x_+\\ x_-\end{pmatrix},\begin{pmatrix} 0&-X^*\\X&0\end{pmatrix}\begin{pmatrix}y_+\\ y_-\end{pmatrix}\right]
+
\fv\left [\begin{pmatrix}x_+\\ x_-\end{pmatrix},\begin{pmatrix}y_+\\ y_-\end{pmatrix}\right]
\\
&=\fa_+[x_+, -X^*y_-]-\fa_-[x_-, Xy_+]+\fa_+[-X^*x_-, y_+]-\fa_-[Xx_+, y_-]
\\&\quad 
+\fv[-X^*x_-, Xy_+]+\fv[Xx_+, -X^*y_-]+\fv[x_+, y_-]+\fv[x_-, y_+]
\\& =\overline{\fa_+[-X^*y_-, x_+]-\fa_-[y_-, Xx_+]+\fv[-X^*y_-, Xx_+]+\fv[ y_-, x_+]}
\\&\quad +
\fa_+[-X^*x_-, y_+]-\fa_-[x_-, Xy_+]+\fv[-X^*x_-, Xy_+]+\fv[x_+, y_-]
\\&=0,
\end{align*} 
 which shows that  \(Y\) is a solution of the block Riccati equation \eqref{RicBlock}. 
\end{proof}
\end{lemma}

\begin{lemma}\label{ricred}  Assume Hypothesis \ref{tech}.
Suppose that \(\fb\) is  a saddle point form with respect to the decomposition \(\cH=\cH_+\oplus \cH_-\) and that \(\cH=\cG_+\oplus \cG_-\).

If  $Y$  solves  to the block form Riccati equation \eqref{RicBlock} and 
$
\cH_A^1\subseteq\Ran(I-Y)|_{\cH_A^1},
$
then  the  decomposition \(\cH=\cG_+\oplus \cG_-\) reduces the form \(\fb\).
\begin{proof}  
Let $Q$ denote the orthogonal projection onto  $\cG(\cH_+,X)$.
Recall that $Q$   is given by the block matrix \eqref{P}
\begin{equation}\label{P**}Q=\begin{pmatrix} (I_{\cH_+}+X^\ast X)^{-1} & (I_{\cH_+}+X^\ast X)^{-1}X^\ast\\
X(I_{\cH_+}+X^\ast X)^{-1}& X(I_{\cH_+}+X^\ast X)^{-1}X^\ast\end{pmatrix}
_{\cH_+\oplus \cH_-}.\end{equation}
 
By hypothesis, one has that $(I-Y)\dom[\fa]\supseteq \dom[\fa]$. Since \(Y\) is a solution of the Riccati equation \eqref{RicBlock}, then necessarily \((I-Y)\dom[\fa]\subseteq\dom[\fa]\). Thus, \(I-Y\) is bijective on \(\dom[\fa]\). So is    the operator
 \begin{equation*}
I-Y^2=(I-Y)J(I-Y)J=\begin{pmatrix}
I_{\cH_+}+X^* X&0\\
0&I_{\cH_-}+X X^\ast
\end{pmatrix},\end{equation*}
where the involution \(J\) is given by \eqref{decJ}.

In particular, the operators 
$I_{\cH_+}+X^* X$ and $I_{\cH_-}+X X^*$ are bijective on $\dom[\fa_+]$ and 
$\dom[\fa_-]$, respectively. Since  $I-Y$ is bijective on $\dom[\fa]$, one also observes that $X$ maps $\dom[\fa_+]$ into  $\dom[\fa_-]$ and 
that $X^*$  maps $\dom[\fa_-]$ into  $\dom[\fa_+]$.  
Taking into account  the explicit representation  \eqref{P**}, one  concludes that the operator \(Q\) maps \(\dom[\fb]=\dom[\fa]\) into itself.

Therefore,  for any  \(\tilde y\in \dom[\fb]\), there exists an \(x\in \dom[\fa_+]\) such that 
$$Q\tilde y= x\oplus Xx.$$
Similarly, for any  \(\tilde x\in \dom[\fb]\) there exists a \(y\in\dom[\fa_-]\) such that
$$Q^\bot \tilde x=-X^\ast y\oplus y.$$

Assuming that  $ x\in \Dom[\fa_+]  $ and $  y\in \Dom[\fa_-]$, we have
\begin{align*}
 \fb[Q^\bot \tilde x,Q \tilde y]&=\fb[-X^\ast y\oplus y,  x\oplus Xx]
 \\&=\fa[-X^\ast y\oplus y,  x\oplus Xx]+\fv[-X^\ast y\oplus y,  x\oplus Xx]
 \\&=\fa_+[-X^\ast y,  x]-\fa_-[y,Xx]+\fv[-X^\ast y, Xx]+\fv[y,x]
\\&= \fa\left [\begin{pmatrix}x\\ 0\end{pmatrix},\begin{pmatrix} 0&-X^*\\X&0\end{pmatrix}
\begin{pmatrix}0\\ y\end{pmatrix}\right ]+
\fa\left [\begin{pmatrix} 0&-X^*\\X&0\end{pmatrix}\begin{pmatrix}x\\ 0\end{pmatrix},\begin{pmatrix}0\\ y\end{pmatrix}\right]
\\&\quad  +
\fv\left [\begin{pmatrix} 0&-X^*\\X&0\end{pmatrix} \begin{pmatrix}x\\ 0\end{pmatrix},\begin{pmatrix} 0&-X^*\\X&0\end{pmatrix}\begin{pmatrix}0\\ y\end{pmatrix}\right]
+
\fv\left [\begin{pmatrix}x\\ 0\end{pmatrix},\begin{pmatrix}0\\ y\end{pmatrix}\right]
\\&=
\fa[f,Yg]+\fa[Yf,g]+\fv[Y f,Yg]+\fv[f,g]\\
&=
0,
\end{align*}
 where we have used the block Riccati equation \eqref{RicBlock}
for $f=x\oplus 0 $ and $   g=0\oplus y$
 on the last step. This implies that $$\fb[Q^\bot \tilde x,Q \tilde y]=0 \quad \text{ for all} \quad \tilde x,\tilde y \in \dom[\fb], 
 $$
 and therefore,  the graph subspace \(\cG_+=\cG(\cH_+,X)\) reduces the form \(\fb\) (see \eqref{rm:2:2}). 
\end{proof}
\end{lemma}

Now we are ready to present  a  generalization of Theorem \ref{vadi2} (i) that yields the block diagonalization of a saddle point form, provided that the latter has a reducing subspace.

\begin{theorem}\label{vadi3}  Assume Hypothesis \ref{tech}.
Suppose that the graph decomposition 
\(\cH=\cG_+ \oplus \cG_-\) reduces the form \(\fb\) and let $U$ be the direct rotation from the subspace $\cH_+$ to the reducing subspace 
$\cG_+$.
Also assume that the decomposition is \(\cH_A^1\)-regular.

Then
$$
\widehat \fb[f,g]=\fb[Uf,Ug], \quad f,g\in \Dom[\widehat \fb]=\Dom[\fb],$$
  is a diagonal form with respect to the decomposition \(\cH=\cH_+\oplus \cH_-\),

\begin{proof}
Due to Theorem \ref{regularSaddlepointForm*}, the Riccati equation  \eqref{RicBlock}  holds if and only if the decomposition  \(\cH=\cG_+ \oplus \cG_-\) reduces the form.  
Then a straightforward computation  shows that 
\begin{equation*}
\fb[(I+Y)f, h]=\fa[f,(I-Y)h]+\fv[Yf,(I-Y)h], \quad f,h \in \dom[\fa].\end{equation*}
Then, taking \(h=(I-Y)^{-1}g\) with \(g\in\dom[\fa]\),  one obtains that 
 \begin{equation}\label{diagvY}
\fb[(I+Y)f, (I-Y)^{-1}g]=\fa[f,g]+\fv[Yf,g],\quad f,g\in \dom[\fa].
\end{equation} 
Since the form  \(\fa\) is diagonal,  and both the form  \(\fv\) and the operator \(Y\) are off-diagonal, it follows that  the form $\fd[f, g]=\fb[(I+Y)f, (I-Y)^{-1}g]$  on $\dom[\fd]= \dom[\fa]$,
is a diagonal form. 

Since \(U=(I+Y)|I+Y|^{-1}=(I-Y)^{-1}|I-Y|\) and \(|I-Y|=|I+Y|\) is a diagonal operator, the equation \eqref{diagvY}  yields
\[\fb[Uf, Ug]=\fa[|I+Y|^{-1}f,|I-Y|g]+\fv[Y|I+Y|^{-1}f, |I-Y|g], \quad f,g\in \dom[\fa],\]
 provided that $|I+Y|=(I-Y^2)^{1/2}$ is bijective on \(\dom[\fa]\). This required bijectivity of $|I+Y|$  follows along similar lines as in  the proof of Lemma 
\ref{Umapsfine}.
\end{proof}
\end{theorem}

It should be noted that the proof of Theorem \ref{vadi3}, compared to the one of Theorem \ref{vadi2} (i), neither requires the  domain stability condition  to hold nor the semi-definiteness of the corresponding reducing graph subspaces $\cG_\pm $. If, however, the domain stability condition holds, the proof of Theorem \ref{vadi2} (i) shows that 
the diagonalization procedure for the unbounded form \(\fb\) in \(\cH\) can be  reduced to the one  of the corresponding bounded self-adjoint operator \(\cB\) in the Sobolev space \(\cH_A^1\) (see \eqref{eq:fbxy}). The form $\widehat \fb$ then splits into the sum of two diagonal forms $\pm \widehat \fb_\pm$,
$$\widehat \fb=\widehat \fb_+\oplus(-\widehat \fb_-),$$that are not necessarily semi-bouded. 
 However, if the saddle point form $\fb$ is {\it a priori}  semi-bounded, the domain stability condition holds automatically and the corresponding diagonal forms $\pm\widehat \fb_\pm$ are semi-bounded and closed. In other words, in this case the statement of  Theorem  \ref{vadi3} can naturally be  extended to the format of Theorem \ref{vadi2}.

\section{Some applications}
In this section, having in mind applications of the developed formalism to the study of  the block Stokes operator from fluid dynamics,  cf.~ \cite{FFMM, Grubb, Reynolds}, 
 we focus on the class of  saddle-point forms provided by Example \ref{illdef}  in the semi-bounded situation. 

We start by  the  following   compactness result that may be of independent interest.

\begin{lemma}\label{lll} Let   $\fb $ be  the   saddle-point form  from  Example \ref{illdef} and $B$ the associated operator. Assume that \(A_+>0\) and that the operator \(A_-\) is bounded and has compact resolvent.
Then the positive spectral subspace of the operator $B$ is a graph subspace,
 $$
      \Ran\left (\EE_B((0, \infty))\right ) =\mathrm{Graph}(\cH_+,X)
      $$ 
			with $X\colon\cH_+\to\cH_-$ a compact contraction. 
 
If, in addition,   \(A_+^{-1}\) is in the Schatten-von Neumann ideal  \(\fS_p\), then \(X\) belongs to \(\fS_{2p}\). 

\begin{proof}

By Theorem \ref{vadi1},
$$
      \Ran(\EE_B((0, \infty)))\oplus (\Ker (B)\cap \cH_+ )=\cG(\cH_+,X),
      $$
with $X$ a contraction.

By  \cite[Theorem 1.3]{SchPaper}  we have that
\[\Ker(B)=(\Ker(A_+)\cap \cK_+)\oplus (\Ker(A_-)\cap \cK_-),\] where 
$$\cK_\pm=\big\{x_\pm\in \cH_\pm\;|\; \fv[x_,x_-]=0 \text{ for all }x_\mp \in \dom[\fa_\mp].\big\}$$
Therefore, if $A_+>0$, then  $\Ker (B)\cap \cH_+=\{0\},$
which proves that 
$$ \Ran(\EE_B((0, \infty)) =\mathrm{Graph}(\cH_+,X).
      $$

Since the reducing subspace $\Ran(\EE_B(0, \infty))$ is  a graph subspace,
the form Riccati equation \eqref{RicBlock} holds. Notice that 
the  Riccati equation \eqref{RicBlock} 
can also be rewritten as the following quadratic equation
\begin{equation}\label{ric}
\fa_+[-X^\ast y,x]-\fa_-[y,Xx]+\fv[-X^\ast y, Xx]+\fv[y,x]=0,\end{equation}
$$ x\in \dom[\fa_+]\subseteq\cH_+,\quad y\in \dom[\fa_-]\subseteq\cH_-,
$$  for  a ``weak solution" $X$.

First, we claim that  
\begin{equation}\label{eq:X}
(I+A_+^{-1/2}X^\ast WA_+^{-1/2})A_+^{1/2}X^\ast=((W+A_-X)A^{-1/2}_+)^*.\end{equation}

Indeed, since  
  \begin{align*}
  \fa_+[-X^\ast y,x]&=-\big\langle A_+^{1/2} {X^\ast} y,A_+^{1/2}x\big\rangle ,
	\\
  \fa_-[y,Xx]&=\langle y, A_-Xx\rangle,
\end{align*}	and  
   $$
\fv[-X^\ast y, Xx]+  \fv[ y,x]= -\langle WX^\ast y, Xx\rangle+\langle y, Wx\rangle,
  $$
  $$ x\in \dom[\fa_+],\quad y\in \dom[\fa_-]=\cH_-,
$$
equation \eqref{ric} can be rewritten as
$$
\langle(A_+^{1/2}+A_+^{-1/2}X^\ast W) {X^\ast} y,A_+^{1/2}x\big\rangle=
\langle   ((W+A_-X)A^{-1/2}_+)^*y,A^{1/2}_+x\rangle, 
$$ 
$$ x\in \dom[\fa_+],\quad y\in \dom[\fa_-]=\cH_-.
$$

Taking into account that  \(A_+^{1/2}\) is a surjective  map from \(\dom[\fa_+]\) onto \(\cH_+\),  we have
\begin{equation*}(I+A_+^{-1/2}X^\ast WA_+^{-1/2})A_+^{1/2}X^\ast=(A_+^{1/2}+A_+^{-1/2}X^\ast W)X^\ast.
\end{equation*}

 Since   $\dom((A_+)^{1/2}) \subseteq\dom(W)$ and $W$ is a closable operator by hypothesis,  the operator \(WA_+^{-1/2}\) is bounded in \(\cH_+\)  (see, e.g., \cite[Problem 5.22]{Kato}).
 In particular,  $(W+A_-X)A^{-1/2}_+$ is bounded and the claim follows taking into account that $$
(A_+^{1/2}+A_+^{-1/2}X^\ast W)X^\ast=((W+A_-X)A^{-1/2}_+)^*.
$$

To complete the proof of the lemma, one observes that 
  $A_++X^\ast W$ is similar to \(\widehat{B}_+\) and since the kernel of \(B\) is trivial,
the kernel of the operator  $A_++X^\ast W$ is trivial as well. Hence, the kernel of the Fredholm operator
\begin{equation*}
F=I+A_+^{-1/2}X^\ast WA_+^{-1/2}\end{equation*} is also trivial
(here we used that  the operator \(WA_+^{-1/2}\) is bounded and that  $A_+^{-1/2}$ is compact). Hence $F$ has a bounded inverse and then, 
from \eqref{eq:X}, we get that 
\begin{equation}\label{bububu}
X^*=A_+^{-1/2}[F^{-1}((W+A_-X)A^{-1/2}_+)^*].
\end{equation}
Since $A_+^{-1/2}$ is compact, it follows that $X^*$ is compact, so is $X$.
From this representation  it also  follows that   \(A_+^{-1/2}\)  and \(X\)  share the same Schatten class membership.
\end{proof}
\end{lemma}

\begin{remark}  Note that in the situation of Lemma \ref{lll}, in the particular case where the off-diagonal part $W$ of the operator matrix \eqref{Bdot} is a bounded operator,  from \eqref{bububu} it also follows (see, e.g., \cite[Satz 3.23]{Wei})
that 
 $X$ belongs to the same Schatten-Von Neumann ideal $\fS_p$ as $A_+^{-1}$ does, cf. \cite[Corollary 2.9.2]{T}.
\end{remark}

As an illustration consider the following example. 
\begin{example}[The Stokes operator revisited]
 Assume  that $\Omega$  is a bounded \(C^2\)-domain in \(\R^d$, $d\geq 2\). In the direct sum of Hilbert spaces 
 $$\cH=\cH_+\oplus\cH_-,$$ where
$\cH_+=L^2(\Omega)^d$ is the ``velocity space" and $\cH_-= L^2(\Omega)$ the ``pressure space",
 introduce the block Stokes operator \(S\)
via the  symmetric sesquilinear form 
\begin{align}\fs[v\oplus p,u\oplus q]&=\nu \langle\text{{\bf grad }}v, \text{{\bf grad} }u\rangle-\vv\langle \div v,q\rangle-\vv\langle p,\div u\rangle\label{formstokes}
\\ &=:\fa_+[v,u]+\fv[v\oplus p,u\oplus q],\nonumber\end{align}
$$\Dom[\fs]=\{v\oplus p\, |\, v\in H_0^1(\Omega)^d, \,\, p\in L^2(\Omega)\}.$$
Here \(\text{{\bf grad}}\) denotes the component-wise application of the standard gradient operator defined  on the Sobolev space \(H_0^1(\Omega)\), with  $\nu>0$ and $\vv\ge 0$  parameters.
\end{example}

It is easy to see that
 the Stokes operator $S$ defined as the self-adjoint operator associated with  the saddle-point form $\fs$,  is
 the Friedrichs extension of 
  the  operator matrix
\begin{equation}\label{StokesMatrix}
\dot S=\begin{pmatrix}-\nu{\bf\Delta}&v^*\grad\\ -v^*\div &0\end{pmatrix}_{\cH_+\oplus \cH_-}
\end{equation}
 defined on 
 $$
\dom(\dot S)=((H^2(\Omega)
\cap  H_0^1(\Omega))^d\oplus H^1_0(\Omega).
 $$
Here  \({\bf\Delta}=\Delta \cdot I_d\) is the vector-valued Dirichlet Laplacian,  with $I_d$  the identity operator  in $\C^d$, 
$\div$   is  the maximal divergence operator from \(\cH_+\) to \(\cH_-\)  on \[\dom(\div)=\{v\in L^2(\Omega)^d\;|\; \div v\in L^2(\Omega)\},\]
 and $(-\grad)$ is its adjoint.

It is  also known  that    
the closure of the operator matrix  \[
{\bf S}= \begin{pmatrix}-\nu{\bf\Delta}&v^*\grad\\ -v^*\div &0\end{pmatrix}_{\cH_+\oplus \cH_-}\] naturally defined on a slightly different domain  
$$
\dom ({\bf S})=(H^2(\Omega)
\cap  H_0^1(\Omega))^d\oplus H^1(\Omega)\supset \dom(\dot S)
$$ is self-adjoint  (see \cite{FFMM}), which yields another characterization for the operator  \(S=S(\nu,\vv)\).

  Clearly  the set 
\(C_0^\infty(\Omega)^d\oplus C_0^\infty(\Omega)\) is a core for the form \(\fs\) and the operator \(S\), so  the form $\fs$ and the Friedrichs extension of the operator matrices $\dot S$ or ${\bf S}$ define the same operator.

 We also remark that  the Stokes operator   is not an off-diagonal operator perturbation of  the  diagonal (unperturbed) operator    \(S(\nu,0)\)
defined on 
$$
\Dom(S(\nu,0))=(H^2(\Omega)
\cap  H_0^1(\Omega))^d\oplus L^2(\Omega)
$$ for the operator matrix \eqref{StokesMatrix} is not a closed operator.

The following proposition can be considered a natural addendum to the known results for the Stokes operator \cite{Atkinson, FFMM, Grubb, Reynolds,LL,  Mennicken:Shkalikov}, see also \cite[Example 2.4.11]{T}.

\begin{proposition}\label{osnt}  Let $\lambda_1(\Omega)$ be the first eigenvalue of the Dirichlet Laplacian on the bounded domain $\Omega\subset \mathbb{R}^d$, $d\ge2$. Then

\begin{itemize}
\item[(i)]
 the positive spectral subspace of the Stokes operator \(S\) can be represented as the graph of a  contractive operator  $X\colon L^2(\Omega)^d\to L^2(\Omega)$
     with
\begin{equation}\label{est}\left\|X\right\|\leq \tan\left(\frac12\arctan\rlad\right)<1,
\end{equation}
where 
\begin{equation}\label{rey}
\rlad=\frac{2\vv}{ \nu\sqrt{ \lambda_1(\Omega)}};
\end{equation}
\item[(ii)] the operator $X$ belongs to the Schatten-von Neumann ideal $\fS_p$ for any $p> d$; 
 \item[(iii)]  the  corresponding direct rotation  $U$  from the ``velocity subspace'' \(L^2(\Omega)^d\) to the positive spectral subspace  of the Stokes operator $S$
 maps the domain of the form onto itself. That is, 
\begin{equation}\label{uuuu}
U\big(H_0^1(\Omega)^d\oplus L^2(\Omega)\big)=H_0^1(\Omega)^d\oplus L^2(\Omega).
\end{equation}

In particular,  the form \eqref{formstokes} and the Stokes operator \(S\) can be block diagonalized
by the unitary transformation $U$.
\end{itemize}

  \begin{proof}  
(i). Due to the embedding
	 $$\dom((-\mathbf{\Delta})^{1/2})=H_0^1(\Omega)^d\subset \{v\in L^2(\Omega)^d\;|\; \div v\in L^2(\Omega)\}=\dom(\div),
	 $$
	 the entries of the operator matrix $\dot S$ satisfy the hypothesis of Example \ref{illdef}, so that the sesquilinear form $\fs$ is a saddle-point form by Lemma \ref{genex2}.
	 The first part of the assertion (i) then follows from  Lemma \ref{lll}.

To complete the proof of (i)
it remains to check the estimate \eqref{est}. 

Recall that if $P$ and $Q$  are orthogonal projections and $\Ran(Q)$ is the  graph  of a bounded operator $X$ from  $\Ran(P)$ to  $\Ran(P^\perp) $,
then  the operator angle $\Theta$  between the subspaces $\Ran(P)$ and $\Ran(Q)$ is  a unique self-adjoint operator in the Hilbert space $\cH$ 
with the spectrum in  $[0,\pi/2]$ 
such that
$$
\sin^2\Theta=PQ^\perp|_{\ran(P)}.
$$
 In this case, 
\begin{equation}\label{tantan}\left\|X\right\|=\tan\left \|\Theta\right\|\end{equation} (see, e.g.,  equation (3.12) in \cite{KMM:1}).

Using the estimate   \cite{Reynolds}
 $$\tan 2 \left\|\Theta\right\|\leq \frac{2\vv}{ \nu\sqrt{ \lambda_1((\Omega)}}
 $$ 
for the operator angle $\Theta$ between the ``velocity subspace''  $\cH_+=L^2(\Omega)^d $ and the positive spectral subspace $\cL_+=\Ran (\EE_S((0, \infty))$ of the Stokes operator,
 one gets the bound \eqref{est} as a consequence of \eqref{tantan}. 
 
(ii). Denote by  $\lambda_k(\Omega)$ the $k^{\text{th}}$-eigenvalue counting multiplicity of the Dirichlet Laplacian on the domain $\Omega\subset \mathbb{R}^d$, $d\ge 2$. By the Weyl's law, the following asymptotics
 $$
  \lambda_k(\Omega)\sim \frac{4\pi^ 2k^{2/d}}{(|B_d||\Omega|)^{2/d}}\quad (\text{as} \quad k\to \infty)
$$ 
holds, see,  e.g., \cite[Theorem 5.1]{BirmanSolomyak}
(here  $|B_d|$ is the volume of the unit ball in $\R^d$ and \(|\Omega|\) is  the volume of the domain \(\Omega\)). Hence, 
the resolvent of the vector-valued Dirichlet  Laplacian $\bf {\Delta}$  
 belongs to the ideal $\fS_p$ for any $p>d/2$. Then, by Lemma \ref{lll},
 we have that 
$$
X\in \fS_p, \quad \text{for any}\quad p>d,
$$ which completes the proof of (ii).

(iii). Since $\fs$ is a semi-bounded saddle-point form, one can apply Theorem  \ref{vadi2} to justify  \eqref{uuuu} as well as the remaining statements of the proposition.
\end{proof}
\end{proposition}

\begin{remark}
The first  part of the assertion (i) is known. It can be verified,  for instance, by
combing  Theorem 2.7.7, Remark 2.7.12 and  Proposition 2.7.13 in \cite{T}.

The generalized Reynolds number  $
\rlad=\frac{2\vv}{ \nu\sqrt{ \lambda_1(\Omega)}}$ given by \eqref{rey}
 has been introduced by Ladyzhenskaya in connection with her analysis of stability of solutions of the \(2D\)-Navier-Stokes equations in bounded domains  \cite{Lad}.
 To the best of our knowledge, the estimate \eqref{est}, the Schatten class  membership $X\in \fS_p$, $p>d$, as well the  mapping property \eqref{uuuu}  of the direct rotation $U$ are new.
 We also note that the diagonalization of $S$ by a similarity transform has already been discussed  and  the one  by a unitary operator has been  indicated, see \cite[Theorem 2.8.1]{T}.
 \end{remark}



\begin{thebibliography}{50}

\bibitem{AL95} V.~M.~ Adamjan, H.~ Langer, \textit{Spectral properties of a class of rational operator valued functions}, J.~ Oper.~ Theory 
\textbf{33} (1995), 259 -- 277.


\bibitem{ALT01} V.~ M.~ Adamyan, H.~ Langer, C.~ Tretter, \textit{Existence and uniqueness
of contractive solutions of some Riccati equations}, J. Funct. Anal. \textbf{
179} (2001), 448 -- 473.


\bibitem{AMM} S.~Albeverio, K.~A.~Makarov, A.~K.~Motovilov,
\textit{Graph subspaces and the spectral shift function}, Canad. J. Math.
\textbf{55} (2003), 449 -- 503.

\bibitem{Atkinson} F.~V.~Atkinson, H.~Langer, R. Mennicken, A.~A.~Shkalikov,
\textit{The essential spectrum of some matrix operators}, Math. Nachr. \textbf{167} (1994), 5 -- 20.

\bibitem{BirmanSolomyak} M.~Sh.~ Birman, M.~Z.~ Solomyak, \textit{The leading term in the asymptotic spectral formula for ``nonsmooth" elliptic problems}, Functional Anal.~ Appl.~ \textbf{4} (1970), 265 -- 275.


\bibitem{Cue12} J.-C.\ Cuenin,
  \emph{Block-diagonalization of operators with gaps, with applications to Dirac operators},
  Rev.\ Math.\ Phys.\ \textbf{24} (2012), 1250021, 31 pp.


\bibitem{Davis} Ch.~Davis, \textit{Separation of two linear subspaces},  Acta Sci. Math. Szeged,  \textbf{19} (1958), 172 -- 187.

\bibitem{Davis:Kahan} Ch.~Davis, W.~M.~Kahan, \textit{The rotation of eigenvectors by a perturbation.
III}, SIAM J. Numer. Anal. \textbf{7} (1970), 1 -- 46.
%
\bibitem{Egorov} A.~I.~Egorov, \textit{Riccati Equations}, 
Russian Academic Monographs, Pensoft, 2007.

\bibitem{FFMM} M.~Faierman, R.~J.~ Fries, R.~Mennicken, M.~M\"oller,
 \textit{ On the essential spectrum of the linearized Navier-Stokes operator},
  Integr. Equ.~ Oper.~ Theory \textbf{38} (2000), 9--27.


\bibitem{Fleige:1} A.~Fleige, \textit{Non-semibounded sesquilinear forms and left-indefinite Srurm-Liouville problems}, Integr.~ Equ.~ Oper.~ Theory \textbf{33} (1999), 20 -- 33.


\bibitem{Fleige:3} A.~Fleige, \textit{Spectral Theory of Indefinite Krein-Feller Differential Operators (Mathematical Research
98)}, Akademie, Berlin, 1996.



\bibitem{FS}D.~ Fuchs, A.~ Schwarz, \textit{Matrix Vieta theorem}, Lie groups and Lie algebras: E. B. Dynkin's Seminar, Amer. Math. Soc. Transl. Ser. 2 \textbf{169}, 15 -- 22, Adv. Math. Sci. \textbf{26}, Amer. Math. Soc., Providence, RI, 1995.

 
\bibitem{G2004} S.~I. ~Gelfand,
\textit{On the  number of solution of a quadratic equation},
in V.~V.~Prasolov, M.~A.~Zfasman (Eds.),  Globus, General Mathematics Seminar, Issue 1, M.: MCCME, pp. 124 -- 133, 2004. (Russian)


\bibitem{GLR} I.~ Gohberg, P.~ Lancaster,  L.~ Rodman, \textit{Invariant Subspaces of Matrices with Applications}, Reprint of the 1986 original, Classics in Applied Mathematics 51, SIAM, Philadelphia, 2006.

\bibitem{GLR2}I.~ Gohberg, P.~ Lancaster, L.~ Rodman, \textit{Matrix Polynomials}, Reprint of the 1982 original, Classics in Applied Mathematics 58, SIAM, Philadelphia, 2009.

\bibitem{Grubb} G.~Grubb, G.~Geymonat, \textit{The essential spectrum of elliptic systems
of mixed order}, Math. Ann. \textbf{227} (1977), 247 -- 276.

\bibitem{pap:1} L.~Grubi\v{s}i\'c, V.~Kostrykin, K.~A.~Makarov, K.~Veseli\'c, \textit{Representation theorems
for indefinite quadratic forms revisited}, Mathematika \textbf{59} (2013), 169 -- 189.

\bibitem{pap:2} L.~Grubi\v{s}i\'c, V.~Kostrykin, K.~A.~Makarov, K.~Veseli\'c, \textit{The $\tan 2\Theta$ for indefinite quadratic forms}, J.~ Spectral Theory \textbf{3} (2013), 83 -- 100.


\bibitem{Reynolds} L.~Grubi\v{s}i\'c, V.~Kostrykin, K.~A.~Makarov, S.Schmitz K.~Veseli\'c, \textit{The Tan \(2\Theta\)-Theorem in fluid dynamics}, arXiv:1708.00509. 


\bibitem{Kato} T.~Kato, \textit{Perturbation Theory for Linear
Operators}, Springer-Verlag, Berlin, 1966.

\bibitem{KMM:1} V.~Kostrykin, K.~A.~Makarov, A.~K.~Motovilov, \textit{Existence and uniqueness of solutions to the
operator Riccati eqution. A geometric approach}, in Yu.~Karpeshina,
G.~Stolz, R.~Weikard, Y.~Zeng (Eds.), \textit{Advances in Differential
Equations and Mathematical Physics}, Contemporary Mathematics \textbf{327},
Amer. Math. Soc., 2003, p. 181 -- 198.

\bibitem{KMM03b}V.~ Kostrykin, K.~A.~ Makarov, A.~ K.~ Motovilov, \textit{On a subspace
perturbation problem}, Proc. Amer. Math. Soc. \textbf{131} (2003), 3469 -- 3476.


\bibitem{KMM:2} V.~Kostrykin, K.~A.~Makarov, A.~K.~Motovilov, \textit{A generalization of the $\tan 2\Theta$
theorem}, in J.~A.~Ball, M.~Klaus, J.~W.~Helton, and L.~Rodman (Eds.),
\textit{Current Trends in Operator Theory and Its Applications}, Operator
Theory: Advances and Applications Vol.~149. Birkh\"{a}user, Basel, 2004, p.~349
-- 372.


\bibitem{KMM05} V. Kostrykin, K.~A.~Makarov, A.~K.~Motovilov, \textit{On the existence of solutions to the operator Riccati equation and the $\tan \theta$ theorem}, Integr.~Equ.~ Oper.~ Theory \textbf{51} (2005), 121 -- 140.

\bibitem{KMM:3} V.~Kostrykin, K.~A.~Makarov, A.~K.~Motovilov,
\textit{Perturbation of spectra and spectral subspaces}, Trans. Amer. Math.
Soc. \textbf{359} (2007), 77 -- 89.

\bibitem{Lad} O.~ A.~Ladyzhenskaya, \textit{The mathematical theory of viscous incompressible flow}, Gordon and Breach Science Publishers, New York, London, Paris, Second Edition, 1969.
%

\bibitem{LR} P.~ Lancaster, L.~ Rodman, \textit{Algebraic Riccati Equations}, Clarendon Press, Oxford, 1995.

\bibitem{LL} L.~D.~ Landau, E.~M.~ Lifschitz, Fluid mechanics. Pergamon Press, New York, Second Edition, 1987.

\bibitem{MSS} K.~A.~Makarov, S.~Schmitz, A.~Seelmann, \textit{On invariant graph subspaces}, Integr. Equ. Oper. Theory \textbf{85} (2016), 399 -- 425.

\bibitem{MSSalt} K.~A.~Makarov, S.~Schmitz, A.~Seelmann, \textit{Reducing graph subspaces and strong solutions to operator Riccati equations}, ArXiv 1307.6439, 2013.


\bibitem{McIntosh:1} A.~McIntosh, \textit{Bilinear forms in Hilbert space},
J. Math. Mech. \textbf{19} (1970), 1027 -- 1045.


\bibitem{Mennicken:Shkalikov} R.~Mennicken, A.~A.~Shkalikov,
\textit{Spectral decomposition of symmetric operator matrices}, Math. Nachr.
\textbf{179} (1996), 259 -- 273.

\bibitem{Plachenov} A.~B.~Plachenov, The matrix equations $\Gamma B\Gamma+\Gamma A-D \Gamma-C=0$, $\Gamma^m+A_1\Gamma^{m-1}+\dots +A_m=0$ and invariant subspaces of block matrices, 2015, preprint.
\bibitem{RS1} M.~Reed, B.~Simon, Methods of modern mathematical physics. I. Functional analysis. Academic Press, Inc., 1980.

\bibitem{RS2} M.~Reed, B.~Simon, Methods of modern mathematical physics. II. Fourier analysis, self-adjointness. Academic Press, Inc., 1975.

\bibitem{SchPaper} S.~Schmitz, \textit{Representation theorems for indefinite quadratic forms without spectral gap}, Integr. Equ. Oper.  Theory \textbf{83} (2015), 73 -- 94.

\bibitem{Sch} K.~Schm\"{u}dgen,
\textit{Unbounded Self-adjoint Operators on Hilbert Space}, Springer, Dordrecht, 2012.

\bibitem{Albrecht} A.~Seelmann, \textit{Notes on the $\sin 2\theta$ Theorem},  Integr.~ Equ. Oper.~ Theory \textbf{79} (2014), 579 -- 597.

\bibitem{T} C.~Tretter, \textit{Spectral theory of block operator matrices and applications}, Imperial College Press, London, 2008.

\bibitem{TW} C.~Tretter, C.~Wyss, \textit{Dichotomous Hamiltonians with unbounded entries and solutions of Riccati equations}, J.~ Evol.~Eq.~ \textbf{14} (2014), 121 -- 153. 

\bibitem{Wei} J.~Weidmann,
\textit{Lineare Operatoren auf Hilbertr\"{a}umen Teil I Grundlagen}, Teubner, Stuttgart, 2000.

\bibitem{JWZ} C.~ Wyss,  B.~ Jacob, H.~ J.~ Zwart, \textit{Hamiltonians and Riccati equations for linear systems with unbounded control and observation operators},  SIAM J. Control Optim. \textbf{50} (2012), 1518 -- 1547.

\bibitem{WW} M.~ Winklmeier, C.~ Wyss, \textit{On the Spectral Decomposition of Dichotomous
and Bisectorial Operators}, Integr.~ Equ.~ Oper.~ Theory \textbf{82} (2015), 119 -- 150.

\end{thebibliography}
\end{document}